\documentclass[aps,pra,twocolumn,groupedaddress,showpacs]{revtex4}
\usepackage[dvips]{graphicx}

\begin{document}

\title{Dark evolution in a time-varying Zeno subspace}

\author{C. K. Law}

\affiliation{Department of Physics, The Chinese University of Hong
Kong, Shatin, Hong Kong SAR, China}


\date{\today}


\begin{abstract}

We investigate the evolution of a quantum system under the
influence of sequential measurements. The measurement scheme
distinguishes whether or not the system is in a specified state
$\left| {f_n} \right\rangle$ at the $n^{\rm th}$ step, where
$\left| {f_n} \right\rangle$ varies with $n$. Dark evolution
corresponds to the situation when all measurements give negative
results. We show that dark evolution is unitary in the continuous
measurement limit. We derive the effective Hamiltonian, and
indicate how $\left| {f_n} \right\rangle$ controls quantum state
transport.

\end{abstract}


\maketitle

\section{Introduction}

The influence of measurements on quantum systems has been an
important subject since the discovery of quantum mechanics.  One
of the most intriguing measurement induced phenomena is quantum
Zeno effect (QZE) in which the time evolution of a system may be
frozen under very frequent observations of the initial state
\cite{van,Sudarshan,peres}. QZE is understood as a consequence of
projection postulate and the quadratic behavior of the survival
probability at short times. Experimental observations of QZE in
atomic systems have been reported \cite{expt1,expt2}. Recently,
Facchi {\it et al.} analyzed the Zeno problem with a more general
approach \cite{F1,F2}. They indicated that quantum evolution can
occur in a restricted Hilbert space (Zeno subspace) defined by
measurement projection operators. Such a Zeno subspace serves as a
basis of useful applications, such as quantum state engineering
\cite{BEC} and decoherence control \cite{rempe,F3}.

QZE has been discussed mainly in situations where same state or
observable is frequently monitored. Since the corresponding
measurement projection operators are constant in time, the
underlying Zeno subspace is stationary. A natural extension is the
inclusion of time varying observations \cite{F2,schulman}. This
involves sequential measurements such that different states are
monitored at different times. Such a time-dependent problem has
been studied in a two-level system \cite{schulman}. Although it
should be expected that time-varying projections would lead to
interesting behavior \cite{ah}, the detailed dynamics has not been
fully explored.

In this paper we investigate this problem in an $N-$level system
($N \ge 3$ in general). The system is subjected to a prescribed
sequence of measurements, such that the $n^{\rm th}$ measurement
detects whether the system is in the state $ \left| {f_n }
\right\rangle$ or not. $ \left| {f_n } \right\rangle$ changes with
$n$, and so the Zeno subspace is time-dependent. The measuring
apparatus is designed such that it can only interact with $ \left|
{f_n } \right\rangle$ at the $n^{\rm th}$ step. Each measurement
simply gives ``Yes'' or ``No'' answer, and it does not provide any
further information about the system. An interesting question is
how the system evolves if all measurements give {\em negative}
results, i.e., ``No'' for all $n$. This corresponds to what we
will call dark evolution in this paper. Such evolution is driven
by measurements, and it occurs even if the Hamiltonian of the
measured system is zero.

Early examples of negative result experiments were discussed by
Renninger \cite{renninger} and Dicke \cite{dicke} who indicated
possible modifications of the measured system if the detector does
not detect anything. Since the state of the detector is not
affected by the measured system, negative result experiments are
sometimes known as interaction-free measurement \cite{vaidman}. In
this regard, the measurement scheme that we will examine is a form
of interaction-free measurement generalized to time-dependent
situations. In order to determine the quantum dynamics, we will
present a Hamiltonian formalism of the problem. In particular, we
will show that dark evolution is unitary and it is governed by an
effective Schr\"odinger equation in the frequent measurement
limit. Some of the main features of quantum states transport will
be discussed.

\section{Dark Evolution}
Let $ \left| {\Psi _n } \right\rangle $ be the state of the system
immediate after the $n^{\rm th}$ measurement. The initial state $
\left| {\Psi _0 } \right\rangle $ is prepared such that it is
orthogonal to $ \left| {f_1 } \right\rangle$. If at any step in
the measurement yields a ``Yes" answer, we have to reset the
system to the initial condition and restart the experiment. This
ensures dark evolution in a single run, and hence the system state
remains pure, assuming decoherence effects are negligible.

Dark evolution is described by the relation $(\hbar =1)$,
\begin{equation}
\left| {\Psi _n } \right\rangle  = \left( {1 - \left| {f_n }
\right\rangle \left\langle {f_n } \right|} \right) e^{-iH \tau}
\left| {\Psi _{n - 1} } \right\rangle
\end{equation}
where $H$ is the Hamiltonian (assumed time-independent) of the
un-measured system, and $\tau$ is the time interval between
measurements. After $M$ measurements, the system state is given
by,
\begin{equation}
\left| {\Psi _M } \right\rangle  = P_M e^{-iH \tau}P_{M - 1}e^{-iH
\tau} \cdot  \cdot \cdot P_2 e^{-iH \tau}P_1e^{-iH \tau} \left|
{\Psi _0 } \right\rangle
\end{equation}
where $P_n  = 1 - \left| {f_n } \right\rangle \left\langle {f_n }
\right|$ is the projection operator. Note that $ \left| {\Psi _M }
\right\rangle $ in Eq. (2) has not been normalized. It is
understood that $|\left\langle {\Psi_M} \right. | \left. {\Psi_M}
\right\rangle |^2$ corresponds to the probability of realizing a
run of the experiment involving $M$ measurements with
negative results.

At time $t=n\tau$, we write $\left| {\Psi (t)} \right\rangle =
\left| {\Psi _n } \right\rangle$ and $ {\left| {f(t )}
\right\rangle }= {\left| {f_n} \right\rangle }$. Assuming $
{\left| {f(t )} \right\rangle }$ is continuous in time, Eq. (1)
gives: $\left| {\Psi (t + \tau )} \right\rangle - \left| {\Psi
(t)} \right\rangle =[ {P(t + \tau )e^{ - iH\tau } - 1} ]\left|
{\Psi (t)} \right\rangle = [ {P(t) + \dot P(t)\tau - i\tau P(t)H -
1} ]\left| {\Psi (t)} \right\rangle + {\cal O}(\tau^2)$, where
$\dot P (t) = d P(t)/dt$ is the time derivative of the projection
operator. In the frequent measurement limit $\tau \to 0$, we have
\begin{equation}
i\frac{d}{{dt}}\left| {\Psi (t)} \right\rangle  = \left[
{P(t)HP(t) + i\dot P(t)} \right]\left| {\Psi (t)} \right\rangle ,
\end{equation}
where the identity $P(t) |\Psi (t) \rangle = |\Psi (t) \rangle$
has been employed. We remark that there are subtle relations
between pulsed observations and continuous observations in
realistic systems \cite{schulman}. Here the $\tau \to 0$ limit is
taken for idealized situations. However, we will show that dark
evolution exists in more general situations (Section III), and
projection measurements are not crucial.

Eq. (3) describes the system evolution under the initial
condition: $\left\langle {f(0)} \right. | \left. {\Psi(0)}
\right\rangle =0$. It is easy to show that $ \frac{{d}}{{dt }}
\langle {f|} \Psi \rangle=-\langle {f|} \dot f \rangle  \langle
{f|} \Psi \rangle$, and so $ \left\langle {f(t)} \right. | \left.
{\Psi(t)} \right\rangle =0$ because of the initial condition.
Therefore $\left| {\Psi (t )} \right\rangle$ remains orthogonal to
$ {\left| {f(t )} \right\rangle }$ at any later time. With this
result, Eq. (3) further gives, $\langle {\Psi (t)|} \dot \Psi (t)
\rangle + \langle {\dot \Psi (t)|}  \Psi (t) \rangle  = 0.$ This
shows that the norm,
\begin{equation}
 \left\langle {\Psi (t) |} \right.\left. \Psi
(t) \right\rangle = 1
\end{equation} is preserved, i.e., dark
evolution is unitary in the frequent measurement limit.

\subsection{Effective Hamiltonians}
To learn how the system evolves for a given $|f(t) \rangle$,
it is useful to cast Eq. (3) in a form of Schr\"odinger equation,
\begin{equation}
i\frac{{d\left| \Psi  \right\rangle }}{{d t}} = H_{D} \left| \Psi
\right\rangle
\end{equation}
where $H_{D}$ ($D$ refers to dark evolution) is an effective
Hamiltonian. We point out that Eq. (3) is {\em not} a
Schr\"odinger equation because the $i \dot P (t)$ term is not
Hermitian. This problem can be overcome by making use of the fact
$\left\langle {f(t)} \right. | \left. {\Psi(t)} \right\rangle =0$
shown above. We can add any term ${\left| {X} \right\rangle
\left\langle {f(t)} \right|}$ (where $\left| {X} \right\rangle$ is
arbitrary) inside the bracket in the right side of Eq. (3) without
changing the evolution of $\left| \Psi \right\rangle$. By choosing
$|X\rangle =2 i | {\dot f(t )} \rangle$, we obtain an effective
Hamiltonian:
\begin{equation}
H_{D}(t)  =P(t)H P(t)+ i\left( {| {\dot f(t )} \rangle \langle
{f(t )} | - | {f(t )} \rangle \langle {\dot f(t )} |} \right)
\end{equation}
which is controlled by $|f(t) \rangle$.

The specification of $|f(t) \rangle$ can be made from the unitary
operator that generates the motion of $|f(t) \rangle$. We assume
that $| {f(t)} \rangle = e^{ - i Kt} | {f(0)} \rangle$, where $K$
is a time-independent Hermitian operator. To see the effects of
$K$, we go to a co-moving frame in which $|f(t) \rangle$ is
stationary. This corresponds to a unitary transformation: $ |
{\tilde \Psi (t)}\rangle = e^{ iKt}| {\Psi (t)} \rangle$. The
corresponding Schr\"odinger equation reads: $i | {\dot {\tilde
\Psi} (t)} \rangle = \tilde H_{D} (t) | {\tilde \Psi (t)}\rangle$,
where the transformed effective Hamiltonian $\tilde H_{D}$ is
given by,
\begin{equation}
\tilde H_{D} (t) = P(0) (e^{ i Kt}He^{ - i Kt}-K)P(0).
\end{equation}
This relation indicates the explicit role of $K$ in the effective
Hamiltonian. In deriving Eq. (7), we have made used of the
relation $ \langle {f(0)}  |  {{\tilde \Psi}(t)} \rangle =0$.

\subsection{Formal solutions}

The formal solution of $|\Psi (t) \rangle$ is given by
\begin{equation}
| {\Psi (t)} \rangle  = e^{ - iKt} {\cal T}\left\{ {\exp \left[ {
- i\int_0^t {dt'\tilde H_D (t')} } \right]} \right\}\left| {\Psi
(0)} \right\rangle ,
\end{equation}
where ${\cal T}$ is the time ordering operator. Further
simplification can be made if $K$ and $H$ commute. In this case
$\tilde H_{D}  = P(0) (H-K)P(0)$ is time-independent, Eq. (8)
becomes,
\begin{equation}
\left| {\Psi (t)} \right\rangle  = e^{ - iKt} e^{-i \tilde H_{D}
t} \left| {\Psi (0)} \right\rangle .
\end{equation}
Note that $K$ and $\tilde H_{D}$ do not commute with each other in
general, we may need to solve the eigenvalue problem:
\begin{equation}
\tilde H_{D} | {u_k } \rangle = \omega_k | {u_k } \rangle
\end{equation}
in order to obtain the explicit form of $| {\Psi (t)} \rangle$.
The general solution can then be expressed in an expansion,
\begin{equation}
| {\Psi (t)} \rangle  = \sum\limits_{k = 1}^{N-1} {c_k e^{ -
i\omega _k t} | {v_k (t)} \rangle }
\end{equation}
where $| {v_k (t)} \rangle \equiv e^{-iKt} | {u_k} \rangle$ and
the coefficients $c_k$ are determined by the initial state.

Eq. (11) reveals the basic structure of the solution when
$[K,H]=0$. Initially, $\left\{ {| {v_k (0)} \rangle} \right\}$
corresponds to the set of eigenvectors of $\tilde H_{D}$. As time
increases, each $| {v_k (t)} \rangle$ evolves unitarily according
to $e^{-iKt}$, the same operator that generates the evolution of
$| {f (t)} \rangle$. Therefore, all $| {v_k (t)} \rangle$ remain
orthogonal to $| {f (t)} \rangle$. These eigenvectors are treated
as a natural set of (time evolving) basis vectors of the system. A
remarkable feature is the emergence  of `new' eigen-frequencies
$\omega_k$ associated with these time varying basis vectors. These
frequencies are neither the eigenvalues of $H$ nor $K$.

We illustrate the intricate coupling between $\omega_k$ and
$|f(t)\rangle$ in a three-level system with $H=0$. Let
$|f(t)\rangle$ be a coherent superposition: $\left| {f(t)}
\right\rangle = a_1 e^{ - i\Omega _1 t} \left| {k_1 }
\right\rangle  + a_2 e^{ - i\Omega _2 t} \left| {k_2 }
\right\rangle  + a_3 e^{ - i\Omega _3 t} \left| {k_3 }
\right\rangle $, were $\Omega_j$ and $|k_j \rangle$ $(j=1,2,3)$
are the eigenvalues and eigenvectors of $K$. In this case, $\tilde
H_D=P(0)KP(0)$ is a $2 \times 2$ matrix, the calculation of its
two eigenvalues, $\omega_{\pm}$, gives $\omega _ \pm   = ( {\xi
\pm \sqrt {\xi ^2 - 4\eta } } )/2$, where $\xi  = \Omega _1 +
\Omega _2  + \Omega _3 - |a_1 |^2 \Omega _1 - |a_2 |^2 \Omega _2 -
|a_3 |^2 \Omega _3$ and $\eta = |a_1 |^2 \Omega _2 \Omega _3  +
|a_2 |^2 \Omega _1 \Omega _3  + |a_3 |^2 \Omega _1 \Omega _2.$ We
see that $\omega_\pm$ depend on $a_j$ and $\Omega_j$ nontrivially.
The situations can become more complicated for higher dimensional
systems.

As a general remark, we note that if $K$ has commensurate
eigenvalues then $|f(t) \rangle$ is cyclic with a certain period
$T$. This means $e^{-iKT}=1$ and hence
\begin{equation}
| {\Psi (T)} \rangle  = \sum\limits_{k = 1}^{N-1} {c_k e^{ -
i\omega _k T} | {v_k (0)} \rangle }
\end{equation}
according to Eq. (11). Since $K$ and $\tilde H_{D}$ generally do
not share the same spectrum, we have $e^{-i\omega_k T} \ne 1$.
Therefore the system in general does not return to the initial
state for a cyclic $|f(t) \rangle$.

\subsection{Quantum state transport}

For the purpose of quantum state transport, a relevent problem is
to find a $\left| {f(t)} \right\rangle$ such that the system
evolves in a prescribed function of time. Such an inverse problem
has a simple solution. It follows from Eq. (3) that $H | {\Psi
(t)} \rangle-i| {\dot \Psi (t )} \rangle$ and $\left| {f(t)}
\right\rangle$ must be parallel. This implies $\left| {f(t)}
\right\rangle$ in the form:
\begin{equation}
\left| {f(t)} \right\rangle  = {\cal N}_f \left( {H | {\Psi (t)}
\rangle-i| {\dot \Psi (t )} \rangle }\right)
\end{equation}
where ${\cal N}_f$ is a factor that can be time-dependent. Such a
factor is determined by the normalization condition: $\langle
{f(t)} | {f(t)} \rangle =1$, which gives
\begin{equation}
{\cal N}_f^{-2} = \langle \dot \Psi (t ) | \dot \Psi (t ) \rangle
+ \langle \Psi (t ) |
 H^2 | \Psi (t ) \rangle.
\end{equation}
Eq. (13) indicates how the measurement state $\left| {f(t)}
\right\rangle$ is designed in order to steer the system state to
evolve in a specified way. However, it is important to remark that
$| {\Psi (t)} \rangle$ cannot be arbitrary because $\langle {\Psi
(t)} | {f(t)} \rangle =0$ must be satisfied. A direct calculation
of the inner product $\langle {\Psi (t)} | {f(t)} \rangle$ in Eq.
(13) leads to the condition:
\begin{equation}
i\langle {\Psi (t)|} \dot \Psi (t) \rangle  = \langle \Psi (t ) |
 H| \Psi (t ) \rangle.
\end{equation}
This is a fundamental restriction that all $| {\Psi(t)} \rangle$
must obey in dark evolution.

Let us discuss $H=0$ systems that highlight the pure influence of
time-varying projective measurements. Physical examples of $H=0$
systems may be found in degenerate Zeeman levels of an atom, and
$| {f (t)} \rangle$ corresponds to a coherent superposition of
these levels. For $H=0$, Eq. (15) implies: $\langle {\Psi (t)|}
\dot \Psi (t) \rangle  = 0$, which corresponds to condition of
{\em parallel transport}. It means that $ |{\Psi (t+ \delta t)}
\rangle$ and $| {\Psi (t)} \rangle$ share the same quantum phase
to first order in $\delta t$, i.e., the local phase change ${{\rm
Arg} [\langle {\Psi (t)|} \Psi (t+\delta t ) \rangle} ] \approx
0$. However, as the system evolves, there is an overall phase
accumulated by the system. Such an accumulated phase is purely
geometrical under the parallel transport condition \cite{phase}.

For $H=0$ systems, Eq. (13) indicates: $\left| {f(t)}
\right\rangle = {\cal N}_f | {\dot \Psi (t )} \rangle$. To provide
an explicit example, suppose $\left| {\Psi(t)} \right\rangle$ is
prescribed by
\begin{equation}
| {\Psi (t )}\rangle  = \sum\limits_{j = 1}^N {\sqrt {p_j } e^{ -
i\nu _j t } | j \rangle }
\end{equation}
where $p_j$ and $\nu_j$ are real constants so that $
\sum\nolimits_{j = 1}^{N} {p_j \nu _j } = 0$ is satisfied for the
parallel transport condition. The required $\left| {f (t )}
\right\rangle$ is given by
\begin{equation}
\left| {f(t)} \right\rangle = {\cal N}_f \sum\limits_{j = 1}^N
{\sqrt {p_j } \nu _j e^{ - i\nu _j t } \left| j \right\rangle }.
\end{equation}

We note that the possibility of steering a ($H=0$) system into an
arbitrary state via suitably designed continuous measurements was
noticed by von Neumann many years ago \cite{van}. This is usually
understood in `bright' measurement configurations, i.e., ``Yes''
detection answers leading to a complete state reduction \cite{eg}.
In contrast, our approach exploits the dark Zeno subspace from
which the detector cannot extract any information (except for
two-level systems in which dark and bright measurements are
equivalent). Finally, we remark that our mechanism of transporting
quantum states should be distinguished from adiabatic passage
\cite{adiabatic}, a technique that is commonly employed for state
preparation. Here dark evolution is guided by projections onto a
Zeno subspace, and adiabatic changes of energy eigenstates are not
necessarily required.

\section{Discussion and summary}
Our formulation so far is based on state projections triggered by
measurements. In essence, dark evolution is due to the existence
of a time varying state $|f(t)\rangle$ that the system cannot
access. As long as $|f(t)\rangle$ can be `simulated' in the
system, dark evolution would occur {\em without} involving any
measurements. One possible mechanism is to shift the energy of
$|f(t)\rangle$ by a large amount relative to the energies of all
other states. Because of energy constraint, a system is forbidden
to reach $|f(t)\rangle$, if the initial state is orthogonal to $|
f(0)\rangle$.

To elaborate the idea, let us consider a system with a model
Hamiltonian,
\begin{equation}
{\cal H} = E  | f(t)\rangle \langle f(t)| .
\end{equation}
By writing the system state vector $|\psi_s (t) \rangle$ as
$|\psi_s (t) \rangle = |\Psi (t) \rangle + \alpha (t) |
f(t)\rangle$, where $|\Psi (t) \rangle$ is orthogonal to $|
f(t)\rangle$, the Schr\"odinger equation ${\cal H} |\psi_s \rangle
= i |\dot\psi_s \rangle$ gives:
\begin{equation}
|\dot \Psi \rangle +  \dot \alpha | f \rangle + \alpha | \dot f
\rangle = -i \alpha E | f \rangle
\end{equation}
and $\alpha (t)$ obeys the equation: $i \dot \alpha = (E-i \langle
f | \dot f\rangle) \alpha - i \langle f | \dot \Psi  \rangle$.
When $E$ is sufficiently large such that $ E \gg \langle f | \dot
\Psi \rangle, \langle f | \dot f  \rangle$, we have $\dot \alpha
(t) \approx 0$ and $\alpha (t) \approx i \langle f(t) | \dot \Psi
(t) \rangle /E$ as an adiabatic solution (where terms with fast
oscillatory phase are neglected). This allows us to recover Eq.
(6) ($H=0$ case) from Eq. (19) by keeping the leading terms and
using $ \langle {f(t)} | {{\dot \Psi}(t)} \rangle = - \langle
{{\dot f}(t)} | { \Psi(t)} \rangle$. The idea of applying a large
coupling term to generate Zeno dynamics was suggested in Ref.
\cite{F3}. The above discussion provides a generalization in
time-varying situations. In particular, we indicate the required
conditions on the large parameter $E$ and the speed of $|f(t)
\rangle$.

To summarize, we show how a nonconstant sequence of projections
would force a measured system to evolve. In particular, we
introduce the notion of dark evolution caused by negative result
measurements in the context of QZE. By varying  $\left| {f(t) }
\right\rangle$ with time, dark evolution enables quantum state
transport under certain basic constraint. Since the state of the
detector is unaffected, quantum coherence of the measured system
is preserved in the Zeno subspace. Our study provides a
Hamiltonian formalism to determine the quantum dynamics in the
continuous measurement limit.

\begin{acknowledgments}
The author acknowledges discussions with M.C. Chu, K.M. Cheung,
and M.F. Cheung. This work is supported in part by the Hong Kong
Research Grants Council (grant No. CUHK4016/03P).
\end{acknowledgments}


\begin{references}
\bibitem{phase} The dynamical phase defined by $
 - i\int_0^t {dt'} \langle {\Psi (t')|}
\dot \Psi (t') \rangle $ is zero for parallel transport.
\bibitem{eg} The simplest example is an arbitrary
rotation of polarization states of a single photon using a series
of polarizers, with the orientations of polarizers change smoothly
along the propagation path of the photon.
\end{references}
\end{document}